\documentclass[10pt]{emulateapj}

\usepackage{amssymb}
\usepackage{amsmath}
\shorttitle{A global model of the UV luminosity and stellar mass functions} 
\shortauthors{Tacchella, Trenti \& Carollo}

\begin{document}

\title{A physical model for the $0\la z\la8$ redshift evolution of the galaxy UV luminosity and stellar mass functions}

\author{Sandro Tacchella\altaffilmark{1},
Michele Trenti\altaffilmark{2} $\&$
C. Marcella Carollo\altaffilmark{1}
}
\altaffiltext{1}{Department of Physics, Institute for Astronomy, ETH Zurich, CH-8093 Zurich, Switzerland}
\altaffiltext{2}{Kavli Institute for Cosmology and Institute of Astronomy, University of  Cambridge, Madingley Road, Cambridge, CB3 0HA, United Kingdom}
\email{tasandro@phys.ethz.ch} 

\slugcomment{{\sc accepted for publication in ApJ Letters:} May 01, 2013}

%-------------------------------------------------%
\begin{abstract} 
  We present a model to understand the redshift evolution of the UV luminosity and stellar mass functions of Lyman Break Galaxies. Our approach is based on the assumption that the luminosity and stellar mass of a galaxy is related to its dark matter halo assembly and gas infall rate. Specifically, galaxies experience a burst of star formation at the halo assembly time, followed by a constant star formation rate, representing a secular star formation activity sustained by steady gas accretion. Star formation from steady gas accretion is the dominant contribution to the galaxy UV luminosity at all redshifts. The model is calibrated by constructing a galaxy luminosity versus halo mass relation at $z=4$ via abundance matching. After this luminosity calibration, the model naturally fits the $z=4$ stellar mass function, and correctly predicts the evolution of both luminosity and stellar mass functions   from $z=0$ to $z=8$.  While the details of star formation efficiency and feedback are hidden within our calibrated luminosity versus halo mass relation, our study highlights that the primary driver of galaxy evolution across cosmic time is the build-up of dark matter halos, without the need to invoke a redshift dependent efficiency in converting gas into stars.  \end{abstract}

\keywords{cosmology: theory --- galaxies: high-redshift  --- stars: formation}

%%%%%%%%%%%%%%%%%%%%%%%%%%%%%%%%%%%%
\section{Introduction} \label{sec:intro}

%%%%%%%%%%%%%%%%%%%%%%%%%%%%%%%%%%%%%%%%%%%%%
\begin{figure*} \begin{center} \leavevmode \includegraphics[scale=0.35]{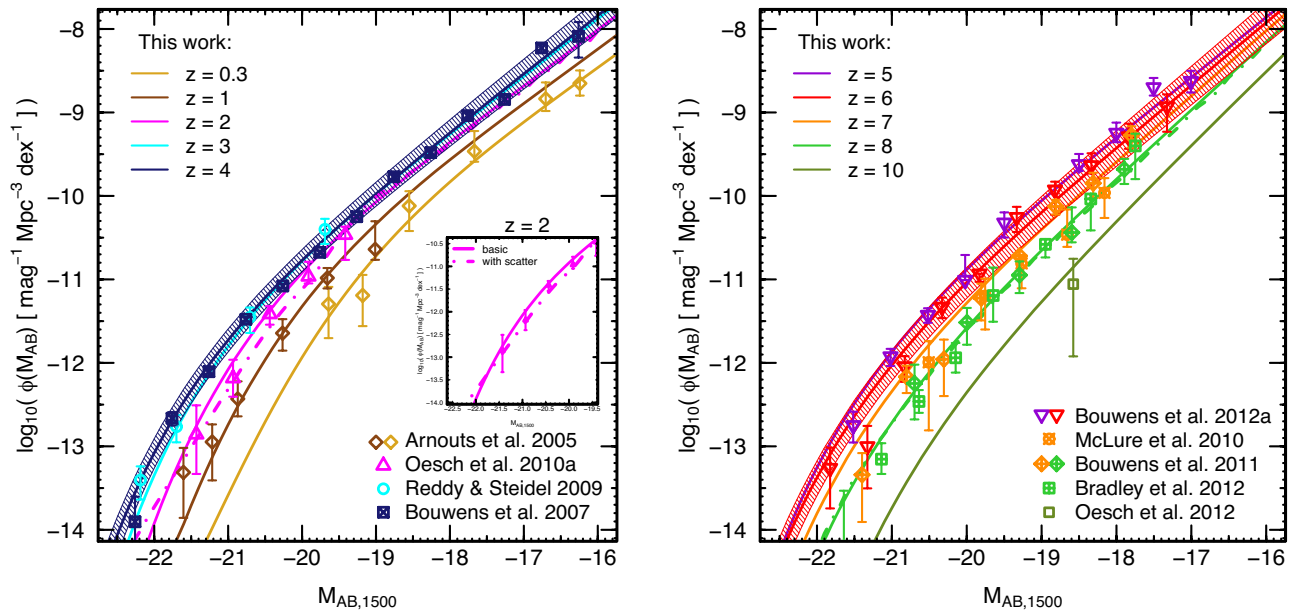} \caption{Observed UV luminosity functions (LFs) at low-$z$ (left) and high-$z$ (right)  (see   Table~\ref{tbl:LFEvolution}).  Shaded areas represent $68\%$ confidence regions at $z=4$ and $z=6$. The dot-dashed lines show the effects on the LFs at $z=2$ and $z=8$ of accounting for the full probability distribution of the halo assembly time, which induces a scatter in the galaxy luminosity versus DM halo mass relation (the inset shows enlargement for $z=2$); this scatter has a negligible influence only at redshifts $z\ga4$. }\label{fig:CompareData}\end{center}\end{figure*}
%%%%%%%%%%%%%%%%%%%%%%%%%%%%%%%%%%%%%%%%%%%%%

The galaxy luminosity function (LF) and the stellar mass function (MF), along with their redshift evolution, summarize key information on galaxy properties and on their evolution with cosmic time. The rest-frame UV $1500~\mathrm{\AA}$ LF in particular can be traced with current technology across the whole redshift range from $z\sim0$ (e.g., \citealt{blanton03,arnouts05,oesch10}) to $z\sim10$ (e.g., \citealt{reddy09,bradley12,oesch12,ellis13}), the current frontier of detection of Lyman Break Galaxy (LBG) populations. Similarly, the stellar MF can be derived from observations in the rest-frame optical \citep{arnouts07,stark13,gonzalez11}. These data enable a self-consistent comparison of star forming galaxies over the entire span of cosmic history. 

A powerful approach to link the properties of galaxies to those of their host dark-matter (DM) halos in a $\Lambda$ Cold Dark Matter ($\Lambda$CDM) cosmology is to use halo occupation distribution models, which give the probability that a halo of mass $M_h$ hosts a galaxy \citep{jing98,peacock00}; these can be generalized into a conditional luminosity function modeling, giving the probability a halo of mass $M_h$ hosts a galaxy with luminosity $L$ \citep{vdbosch03,cooray05a}. This approach provides a $L(M_h)$ relation between galaxy luminosity and DM halo mass, derived at each redshift through ``abundance matching'' \citep[e.g.,][]{mo96,vale04}, which generally includes a duty cycle parameter so as to populate with UV luminous galaxies only a fraction of DM halos \citep{cooray05a,lee09}. It is quite successful in providing a description of the LF, but it does not provide a physical explanation for it. 

Our approach aims at identifying the key drivers of the evolution of galaxy properties with the least amount of assumptions. We showed in \citet{trenti10} that the LF at $z\gtrsim5$ is successfully modeled by assuming that UV bright galaxies are present, at any cosmic epoch, only in halos assembled within $\Delta t$ ($\Delta t\sim10^8~\mathrm{Myr}$). This results in a duty-cycle which is physically motivated, defined without free parameters, and dependent on redshift and halo mass. While this model well reproduces the observed rest-frame UV LF evolution at very high redshifts, it cannot be extrapolated down to redshifts $z\la4$, since at such late epochs DM halos older than a few $10^8~\mathrm{yr}$ are likely to host UV bright galaxies.

In this Letter we expand the \citet{trenti10} model by making the more realistic assumption that, at any epoch, all massive DM halos host a galaxy with a star formation history (SFH) that is related to the time of halo assembly. Note that the duty cycle inserted by \citet{trenti10} is not necessary in our model, which adopts a physical prescription to connect the UV luminosity to a given halo. We anchor our model to the observed LF at $z=4$, and evolve it towards higher ($z\approx8$) and lower ($z\approx0$) redshifts with a simple physical prescription that enables us to explore the origin of the observed UV LF evolution. Our new model features $(i)$ a burst of star formation at halo assembly time, followed by (ii) constant star formation with rate inversely proportional to the halo assembly time (halos at a given mass and different redshift accrete the same gas but over a different timescale). These assumptions, calibrated at $z=4$, are able to reproduce the evolution of the LF, cosmic mass density ($\rho_{M_{\star}}$), and specific star formation rate (sSFR) across 13 billion years of cosmic time. This good match between model and observations is achieved with a dominant contribution to the UV luminosity, at all epochs, of a continuous mode of star formation, fueled by gas accretion.

This Letter is organized as follows. Section~\ref{sec:Obs} summarizes the observational datasets we aim at modeling. Section~\ref{sec:model} describes the model and its calibration.  Section~\ref{sec:Res} presents our results and discusses model uncertainty.  Section~\ref{sec:Con} highlights some concluding remarks. We adopt WMAP5 cosmology: $\Omega_{\Lambda,0}=0.72$, $\Omega_{m,0}=0.28$, $\Omega_{b,0}=0.0462$, $\sigma_8=0.817$, $n_s=0.96$, and $h=0.7$ \citep{komatsu09}.

%%%%%%%%%%%%%%%%%%%%%%%%%%%%%%%%%%%%
\section{Observational Data}\label{sec:Obs}

%%%%%%%%%%%%%%%%%%%%%%%%%%%%%%%%%%%%%%%%%%%%%
\begin{figure} \centering \includegraphics[scale=0.34]{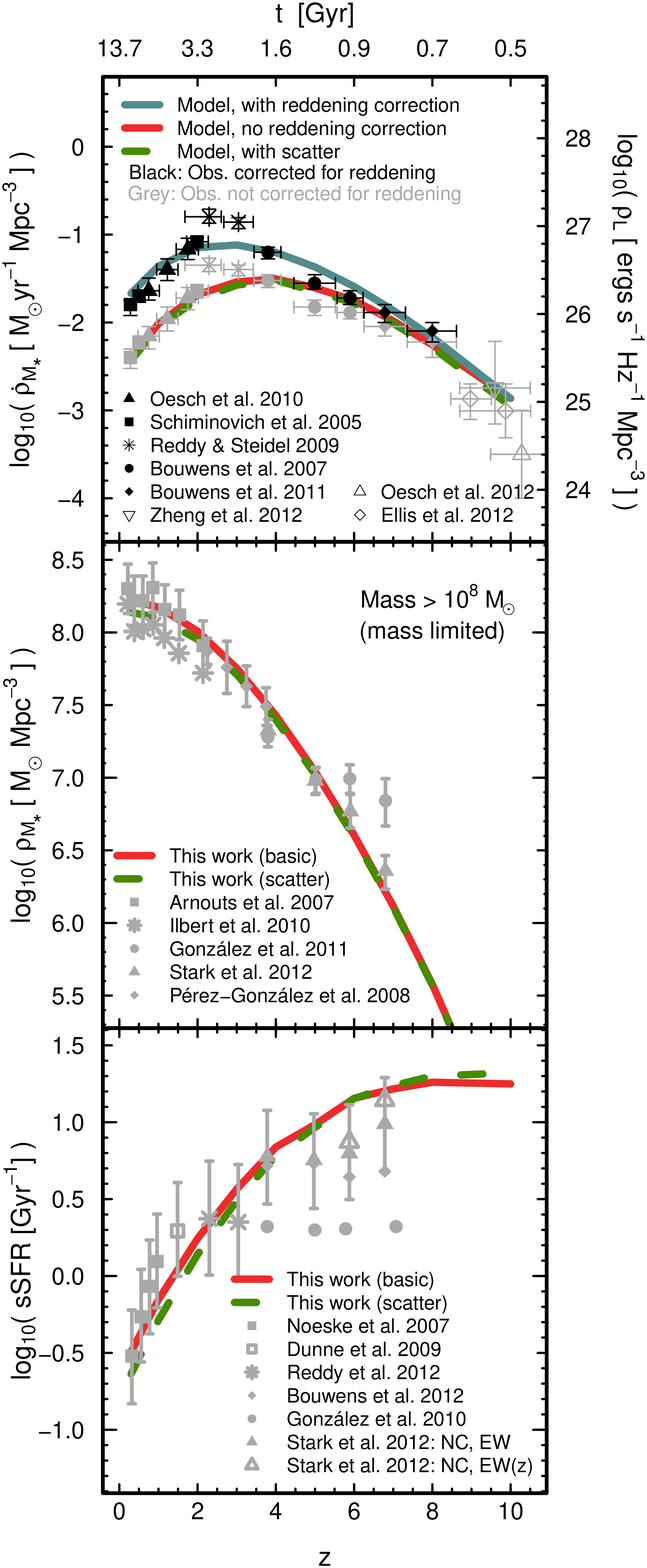} \caption{Top panel: Star formation rate density ($\dot{\rho}_{M_{\star}}$) and luminosity density ($\rho_L$) derived by integrating the model UV LFs in comparison with the observations before (gray points) and after dust correction (black points). Both $\rho_L$ and $\dot{\rho}_{M_{\star}}$ are given for $L\geq0.05~L^{\ast}_{z=3}$ ($M_{AB}\leq-17.7$). Middle: Evolution in the stellar mass density ($\rho_{M_{\star}}$), computed by integrating the stellar mass function to a fixed stellar mass limit of $10^8~\mathrm{M_{\sun}}$. Bottom: Evolution in the sSFR as a function of redshift for a galaxy with $M_{\ast}=5\times10^9~\mathrm{M_{\sun}}$. The red lines show our standard model predictions, while the dashed green lines show the model that includes the full probability distribution of the halo assembly time (i.e. with scatter in the $L(M_h,z)$ relation).}\label{fig:SFRD}\end{figure}
%%%%%%%%%%%%%%%%%%%%%%%%%%%%%%%%%%%%%%%%%%%%%

Figure~\ref{fig:CompareData} and Table~\ref{tbl:LFEvolution} summarize the literature compilation of observed UV LFs we use, including the parameters for the best-fit Schechter LF functions ($\phi(L)=\phi^*(L/L^*)^{\alpha}\exp{-(L/L^*)}$). Observed stellar mass densities ($\rho_{M_*}$) integrated above $M_{*,min}=10^8~\mathrm{M_{\sun}}$ and sSFR at $M_{\star}=5\times10^9~\mathrm{M_{\sun}}$ are shown in Figure~\ref{fig:SFRD}. The figure also includes the cosmic star formation rate density $\dot{\rho}_{M_{\star}}$, obtained by converting, using the \citet{madau98} relation, the luminosity density $\rho_L$ integrated to $L_{min}=0.05~L^{\ast}_{z=3}$ (corresponding to $M_{AB}=-17.7$). Over plotted to observations are our model predictions, obtained with the prescriptions described below.

%%%%%%%%%%%%%%%%%%%%%%%%%%%%%%%%%%%%
\section{Model Description}\label{sec:model}

Our model links the UV LF to abundance of DM halos at the same epoch, from $z=0$ to $z\sim10$, adopting a physical recipe for star formation with dependence on halo assembly time.

\subsection{Halo Assembly Time}\label{subsec:AssemTime}

We adopt the halo assembly time (redshift) as defined by \citet{lacey93} as typical timescale for galaxy formation. The assembly redshift $z_a$ of a halo of mass $M_h$ at redshift $z$ is the redshift at which the mass of the main progenitor is $M_h/2$, which can be calculated within the extended Press-Schechter formalism \citep{bond91}. For this, we use the ellipsoidal collapse model \citep{sheth01}, which reproduces well numerical simulation results \citep{giocoli07}. We adopt, as the fiducial assembly time for each halo, the median of the probability distribution associated with each halo (shown in the top-left panel of Figure~\ref{fig:model} for different redshifts),  but we also account for the full probability distribution of $z_a$ to compute the scatter in the $L(M_h,z)$ relation and validate our simpler assumption of a median value for $z_a$. At a given mass, halos are assembled faster at higher $z$, with important consequences on the UV properties of stellar populations.

\subsection{Star Formation Modeling} \label{subsec:GalLum}

We populate halos with stars based on the Simple Stellar Population (SSP) models of \citet{bruzual03}, adopting a Salpeter IMF \citep{salpeter55} between $M_L=0.1~\mathrm{M_{\sun}}$ and $M_U=100~\mathrm{M_{\sun}}$. We use constant stellar metallicity $Z=0.02~\mathrm{Z_{\sun}}$, neglecting redshift evolution as there is little dependence of the UV luminosity on metallicity. We define as $l(t)$ the resulting luminosity at $1500$~\AA~for a SSP of age $t$ and stellar mass $1~\mathrm{M_{\sun}}$.

For a halo at a given redshift, we set  the start of the SFH to coincide with  the halo assembly time $t_H(z_a)$. Specifically, we parametrize the SFHs through  a short-duration burst at the halo assembly time, followed by a constant SFR period. This latter term is normalized by $1/t_{age}$, with $t_{age}=t_H(z)-t_H(z_a)$ and $t_H(z)$ the age of the Universe at redshift $z$. The  ``burst mode''  integrates the  total stellar mass  produced from the earliest epochs down to $t_{H}(z_a)$. This integrated contribution of early star formation activity makes up for about half of the stellar mass; as $t_{H}(z_a)\ga10^8$ yr, the burst adds up however relatively little, i.e., $\la20\%$, to the UV LF (see Section~\ref{sec:Res}). During the continuous ``accretion mode'', halos accrete  of order $M_h/2$ within the $t_{age}$ time scale. Since $t_{age}$ depends strongly on $z$ (see Figure~\ref{fig:model} top left panel), the accretion rate changes as well: halos of the same mass at higher redshifts have naturally higher accretion rates, as also indicated by other studies \citep{genel08,dekel09}. The resulting halo luminosity is:
%%%%%%%%
\begin{equation}\label{eq:SFH}\begin{split}
L(M_h,z) & =x\cdot\left[\eta(M_h)M_hl(t_{age})\right] \\ 
& +(1-x)\cdot\left[\varepsilon(M_h)M_h\frac{1}{t_{age}}\int_0^{t_{age}}l(t)dt\right],
\end{split}\end{equation}
%%%%%%%%
The first term $\eta(M_h)$ describes the efficiency of the burst episode, while the second term $\varepsilon(M_h)$ describes the efficiency of the accretion mode. Both efficiencies are assumed to be redshift independent (see \citet{behroozi13}). The free parameter $x$ controls the relative contribution of the initial burst to the total luminosity of the galaxy at $z=4$.

From Equation~\ref{eq:SFH}, the resulting stellar mass, i.e., the time integral of the star formation rate is:
%%%%%%%%
\begin{equation}\label{eq:StellarMass}
M_{\star}(M_h)=x\cdot\eta(M_h)M_h+(1-x)\cdot\varepsilon(M_h)M_h,
\end{equation}
%%%%%%%%
which can be used to obtain stellar densities and specific star formation rates.

\subsection{Dust Extinction}\label{subsec:Dust}

Dust extinction significantly affects the observed UV flux, especially at $z\lesssim4$ (see Figure~\ref{fig:SFRD}, top panel).   Following \citet{smit12}, for a spectrum modeled as $f_{\lambda}\sim\lambda^{\beta}$, we assume a linear relation between the UV-continuum slope $\beta$ and luminosity ($<\beta>=\frac{d\beta}{dM_{\mathrm{UV}}}(M_{\mathrm{UV,AB}}+19.5)+\beta_{M_{\mathrm{UV,AB}}}$). Assuming a dependence of UV extinction on $\beta$ as $A_{\mathrm{UV}}=4.43+1.99\beta$ \citep{meurer99}, and a Gaussian distribution for $\beta$ at each $M_{\mathrm{UV}}$ value (with dispersion $\sigma_{\beta}=0.34$),   the average $<A_{\mathrm{UV}}>$ is given by $<A_{\mathrm{M_{UV}}}>=4.43+0.79\ln(10)\sigma_{\beta}^2+1.99<\beta>$. We adopted the value of $0$ for any negative $<A_{\mathrm{UV}}>$. Values for $\frac{d\beta}{dM_{\mathrm{UV}}}$ and $\beta_{M_{\mathrm{UV,AB}}}$ are taken from Table~5 of \citet{bouwens12a} and are listed in Table~\ref{tbl:LFEvolution}. We extrapolated $\beta_{M_{\mathrm{UV,AB}}}$ to higher and lower redshifts, while letting $\frac{d\beta}{dM_{\mathrm{UV}}}$ constant at the $z = 4$ value, since uncertainties in this latter parameter are large.

%%%%%%%%%%%%%%%%%%%%%%%%%%%%%%%%%%%%%%%%%%%%%
\begin{deluxetable*}{ccccccccccccc}
\leavevmode
\tablecolumns{13}
\tablewidth{0pc}
\tablecaption{Best Fit Schechter Function Parameters for our Model UV LFs and Observed LFs (from the Literature). \label{tbl:LFEvolution}}
\tablehead{
\colhead{}&\colhead{}&\multicolumn{3}{c}{Model Prediction}&\colhead{}&\multicolumn{4}{c}{Observed LF}&\colhead{}&\multicolumn{2}{c}{UV continuum parameters}\\
\cline{3-5}\cline{7-10}\cline{12-13}\\
\colhead{Redshift} & \colhead{} & \colhead{$(\phi^{\ast})_{{-3}^a}$}  & \colhead{$M^{\ast}$} & \colhead{$\alpha$}  &  \colhead{}  & \colhead{$(\phi^{\ast})_{{-3}^a}$}  & \colhead{$M^{\ast}$} & \colhead{$\alpha$} & \colhead{Ref} & \colhead{} & \colhead{$\beta_{\mathrm{M_{UV}=-19.5}}$}  & \colhead{$\frac{d\beta}{dM_{\mathrm{UV}}}$} 
}
\startdata
$z=0.3$ 	& & $4.2\pm0.1$ & $-18.9\pm0.1$ & $-1.29\pm0.05$ 
					& & $6.2\pm1.8$ & $-18.4\pm0.3$ & $-1.19\pm0.15$ & (1) 
					& & $-1.45$ & $-0.13$ \\
$z=1$  		& & $1.6^{+0.2}_{-0.1}$ & $-19.9\pm0.1$ & $-1.63^{+0.04}_{-0.02}$ 
					& & $1.1\pm0.8$ & $-20.1\pm0.5$ & $-1.63\pm0.45$ & (1) 
					& & $-1.55$ & $-0.13$ \\
$z=2$  		& & $2.2^{+0.2}_{-0.1}$ & $-20.3^{+0.2}_{-0.1}$ & $-1.60^{+0.04}_{-0.06}$ 
					& & $2.2\pm1.8$ & $-20.2\pm0.5$ & $-1.60\pm0.51$ & (2) 
					& & $-1.70$ & $-0.13$ \\
$z=3$  		& & $1.72\pm0.01$ & $-20.9^{+0.3}_{-0.1}$ & $-1.68^{+0.05}_{-0.07}$ 
					& & $1.7\pm0.5$ & $-21.0\pm0.1$ & $-1.73\pm0.13$ & (3) 
					& & $-1.85$ & $-0.13$ \\
$z=4$  		& & $1.30\pm0.01$ & $-21.0^{+0.2}_{-0.3}$ & $-1.73^{+0.07}_{-0.05}$ 
					& & $1.3\pm0.2$ & $-21.0\pm0.1$ & $-1.73\pm0.05$ & (4) 
					& &  $-2.00$ & $-0.13$ \\
$z=5$  		& & $1.4\pm0.1$ & $-20.6^{+0.2}_{-0.3}$ & $-1.77^{+0.11}_{-0.05}$ 
					& & $1.4^{+0.7}_{-0.5}$ & $-20.6\pm0.2$ & $-1.79\pm0.12$ & (5) 
					& &  $-2.08$ & $-0.16$ \\
$z=6$  		& & $1.4\pm0.1$ & $-20.4^{+0.4}_{-0.2}$ & $-1.76^{+0.14}_{-0.12}$ 
					& & $1.4^{+1.1}_{-0.6}$ & $-20.4\pm0.3$ & $-1.73\pm0.20$ & (5) 
					& &  $-2.20$ & $-0.17$ \\
$z=7$  		& & $0.9\pm0.1$ & $-20.2\pm0.2$ & $-1.84^{+0.12}_{-0.17}$ 
					& & $0.9^{+0.7}_{-0.4}$ & $-20.1\pm0.3$ & $-2.01\pm0.21$ & (6) 
					& &  $-2.27$ & $-0.21$ \\
$z=8$  		& & $0.5\pm0.1$ & $-20.2^{+0.4}_{-0.2}$ & $-1.92^{+0.11}_{-0.15}$ 
					& & $0.4^{+0.4}_{-0.2}$ & $-20.3^{+0.3}_{-0.3}$ & $-1.98^{+0.2}_{-0.2}$ & (7) 
					& & $-2.34$ & $-0.25$ \\
$z=10$		& & $0.2\pm0.1$ & $-19.74^{+0.3}_{-0.5}$ & $-2.18^{+0.25}_{-0.02}$ 
					& & $0.1\pm0.1 $ & $-19.6^b$ & $-1.73^b$ & (8) 
					& & ---$^c$ & ---$^c$ \\
\enddata
\tablecomments{The best fit Schechter Function parameters, as a function of redshift (Column 1), for our model-predicted LFs (Columns 2-4; see Figure~\ref{fig:CompareData}),  and for observed UV LFs (taken from the literature; Columns 5-8). Quoted errors for our model predictions are derived by propagating the uncertainty in the $L(M_h,z=4)$ calibration (see also Figure~\ref{fig:model}). The last two columns show the adopted slopes $\beta_{\mathrm{M_{UV}=-19.5}}$ and intercepts $\frac{d\beta}{dM_{\mathrm{UV}}}$ to the UV-continuum slope $\beta$ to UV luminosity relationship as in  \citet[][Table 5]{bouwens12a}. }
\tablenotetext{a}{Units: $10^{-3}\mathrm{Mpc}^{-3}$.}
\tablenotetext{b}{Values were fixed for the fit of the Schechter function.}
\tablenotetext{c}{No dust correction at $z \sim 10$.}
\tablerefs{ (1) \citet{arnouts05}; (2) \citet{oesch10}; (3) \citet{reddy09}; (4) \citet{bouwens07}; (5) \citet{bouwens12b}; (6) \citet{bouwens11}; (7) \citet{bradley12}; (8) \citet{oesch12}. }
\end{deluxetable*}
%%%%%%%%%%%%%%%%%%%%%%%%%%%%%%%%%%%%%%%%%%%%%

\subsection{Model Calibration}\label{subsec:Derive_LMh}

\begin{figure*} \begin{center} \leavevmode \includegraphics[scale=0.28]{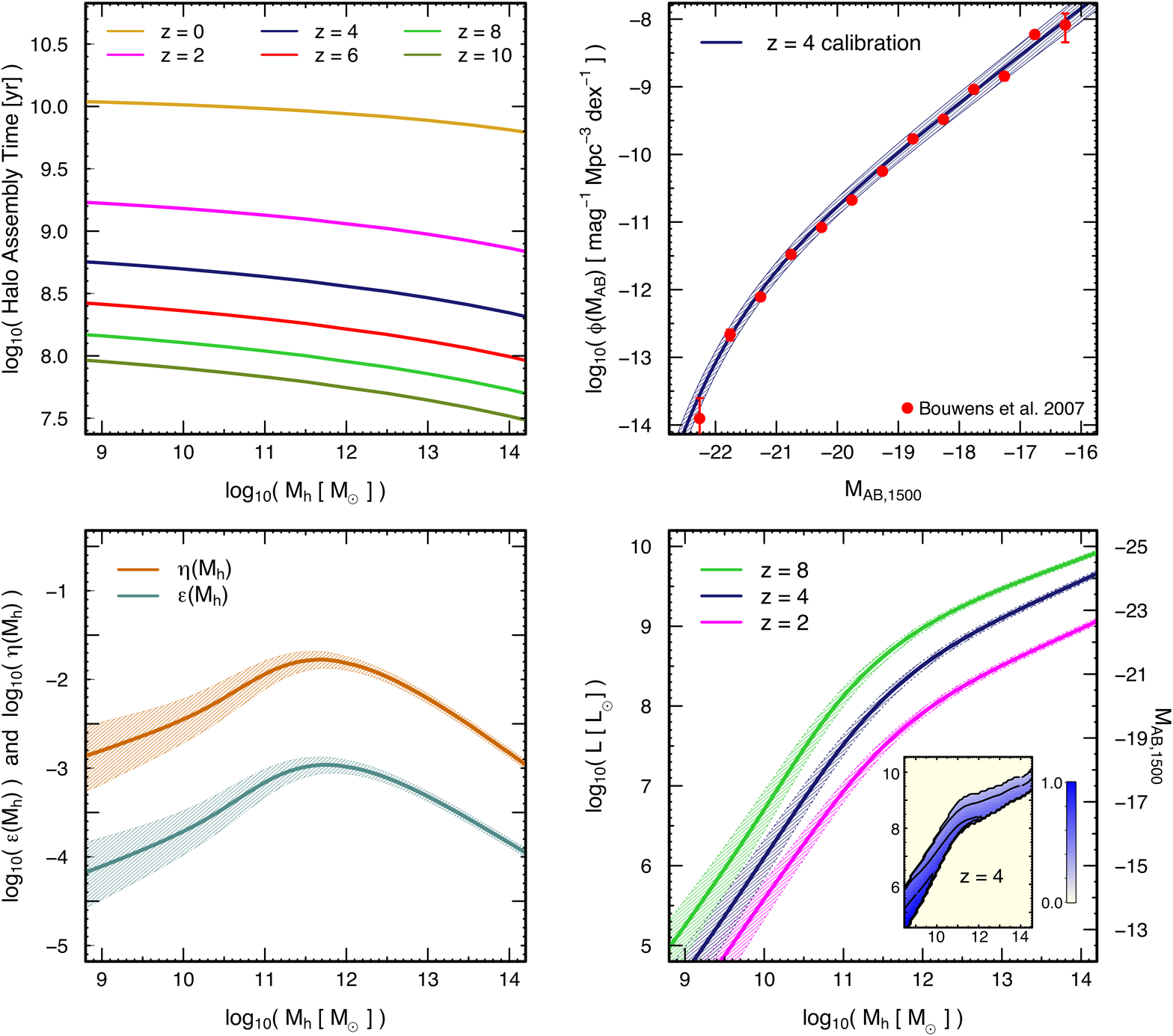}
\caption{Upper-left panel: DM halo assembly time $t_{age}$ versus halo mass and redshift. Upper-right panel: Calibration of our model using the  \citet{bouwens07} LF  at $z=4$. Bottom-left panel: Star formation efficiencies $\eta (M_h)$ and $\varepsilon (M_h)$ for the burst and  accretion mode, respectively. The highest efficiencies are in the range $M_h\sim10^{11}-10^{12}~\mathrm{\mathrm{M_{\sun}}}$ . Bottom-right panel: The relation between galaxy luminosity and DM halo mass, $L(M_h,z)$, plotted at $z=2$, 4, and 8. In all panels, shaded areas represent $68\%$ confidence regions. The inset shows the scatter in the $L(M_h,z)$ relation at $z=4$ produced by sampling the whole probability distribution of halo assembly times.}\label{fig:model}
\end{center}\end{figure*}

To calibrate $\eta(M_h)$ and $\varepsilon(M_h)$ we perform abundance matching at $z=4$, assuming one galaxy per halo and equating the number of galaxies with luminosity greater than $L$ (after dust correction) to the number of halos with mass greater than $M_h$:
%%%%%%%%
\begin{equation}\label{eq:AbMatch}
\int_{M_h}^{+\infty}n(\tilde{M_h},z=4)d\tilde{M_h}=\int_L^{+\infty}\phi(\tilde{L},z=4)d\tilde{L},
\end{equation}
%%%%%%%%
where $n(M_h, z)$ is the MF of DM halos obtained adopting \citet{sheth99} MF. This gives us a luminosity versus halo mass relation at $z=4$, $L(M_h, z=4)$, shown in the bottom-right panel of Figure~\ref{fig:model}. From this we can then infer $\eta (M_h)$ and $\varepsilon (M_h)$ by solving Equation~\ref{eq:SFH} (bottom-left panel of Figure~\ref{fig:model}). We calibrate these two quantities independently, so that their linear combination also satisfies $L(M_h, z=4)$ by construction. The shaded areas in both panels represent the uncertainty in the model calibration derived by varying the $z=4$ LF parameters within the $1\sigma$ confidence regions in Figure 3 of \citet{bouwens07}. From the bottom-left panel of Figure~\ref{fig:model} it is immediate to see that halos with $M_h\sim10^{11}-10^{12}~\mathrm{M_{\sun}}$ have the highest specific star formation efficiencies. This is not surprising, given the shapes of the LFs and DM MF.

Our final calibration step is selecting a value for the only free parameter in the model, $x$, i.e., the contribution of the burst to the total luminosity at $z=4$. For this we compute model predictions over the redshift range $0\la z\la8$ with varying $x$ values, and adopt  the value of $x$ which minimizes the residuals relative to the observed LFs. The best match to observations is given by  $x=0.1$, which is a $10\%$ of contribution from the initial burst to the total halo luminosity at $z=4$; the model is however not very sensitive to the exact value for as long as $x\ll1$ (see top panels of Figure~\ref{fig:sensitivity}). 

%%%%%%%%%%%%%%%%%%%%%%%%%%%%%%%%%%%%%%%%%%%%%
%%%%%%%%%%%%%%%%%%%%%%%%%%%%%%%%%%%%%%%%%%%%%

%%%%%%%%%%%%%%%%%%%%%%%%%%%%%%%%%
\section{Results and Discussion} \label{sec:Res}

The curves in the bottom-right panel of Figure~\ref{fig:model} show our predictions for the observed $L(M_h,z)$ relation at different redshifts. A decreasing contribution from dust is the main cause for the brightening of the relation towards higher redshifts at fixed halo mass. Note that these assume a single assembly time for a given halo (see Section~\ref{subsec:AssemTime}). Taking into account the whole probability distribution for the halo assembly time leads to scatter in the $L(M_h,z)$ relation (shown in the inset of Figure~\ref{fig:model}, bottom-right panel), but such model has overall similar predictions in terms of the observed LF as shown in Figure~\ref{fig:CompareData} for $z=2$ and $z=8$. Therefore, we focus primarily on our canonical model without scatter.

The  predictions for the LFs over the $z\la10$ time span are shown overplotted to the observations in Figure~\ref{fig:CompareData}. In addition, the model reproduces both low-$z$ and high-$z$ UV LFs remarkably well, suggesting that the evolution of the UV LF across most of cosmic time  can indeed result from the redshift evolution of the halo MF, coupled with simple star formation histories beginning at the halo assembly time. 

The model, calibrated to the observed Schechter LF at $z=4$, produces Schechter functions at all other epochs. Furthermore, the predicted LFs well approximates the Schechter functions with the observed best-fit parameters reported in Table~\ref{tbl:LFEvolution} (Figure~\ref{fig:CompareData}).  In particular, the model correctly describes the evolution of the faint-end slope $\alpha$, from its shallow low-$z$ value $\alpha(z\approx0)\sim-1.3$ to the steepening observed  at $z\gtrsim6$, where $\alpha\la-1.7$. At high-$z$, the model  LFs are similar to the LFs predicted by \citet{trenti10}; in addition, our new model is also  successful in  reproducing the observed LFs also at low redshifts, all the way down to $z\simeq0$, resolving the puzzling quick rise of $\alpha$ from $z\simeq0$ to 1 \citep[e.g., ][]{oesch10}.  

Figure~\ref{fig:SFRD} shows the model predictions for the redshift evolution of the star formation rate density $\dot{\rho}_{M_{\star}}(z)$ and luminosity density $\rho_L$($z$) (top panel). The model-observation agreement for $\dot{\rho}_{M_{\star}}(z)$ is again very good at all epochs from $z\approx0$ to $z\approx8$. At $z\sim10$ the model appears to over-predict $\dot{\rho}_{M_{\star}}$ as measured by \citet{oesch12} by $\sim0.8$ dex. This might be due to very short assembly times for $z\sim10$ halos ($t_{age}\lesssim10^8~\mathrm{Myr}$), hence to a dominant contribution to the UV light from the very young stellar populations produced in the burst mode. Another possibility is sample variance in the observations. In fact, \citealt{zheng12} derive from a gravitationally lensed source in CLASH $\dot{\rho}_{M_{\star}}(z=10)=(1.8^{+4.3}_{-1.1})\times10^{-3}~\mathrm{M_{\sun}Mpc^{-3}yr^{-1}}$, in agreement with our model predictions.

%%%%%%%%%%%%%%%%%%%%%%%%%%%%%%%%%%%%%%%%%%%%%
\begin{figure*}  \begin{center} \leavevmode \includegraphics[scale=0.25]{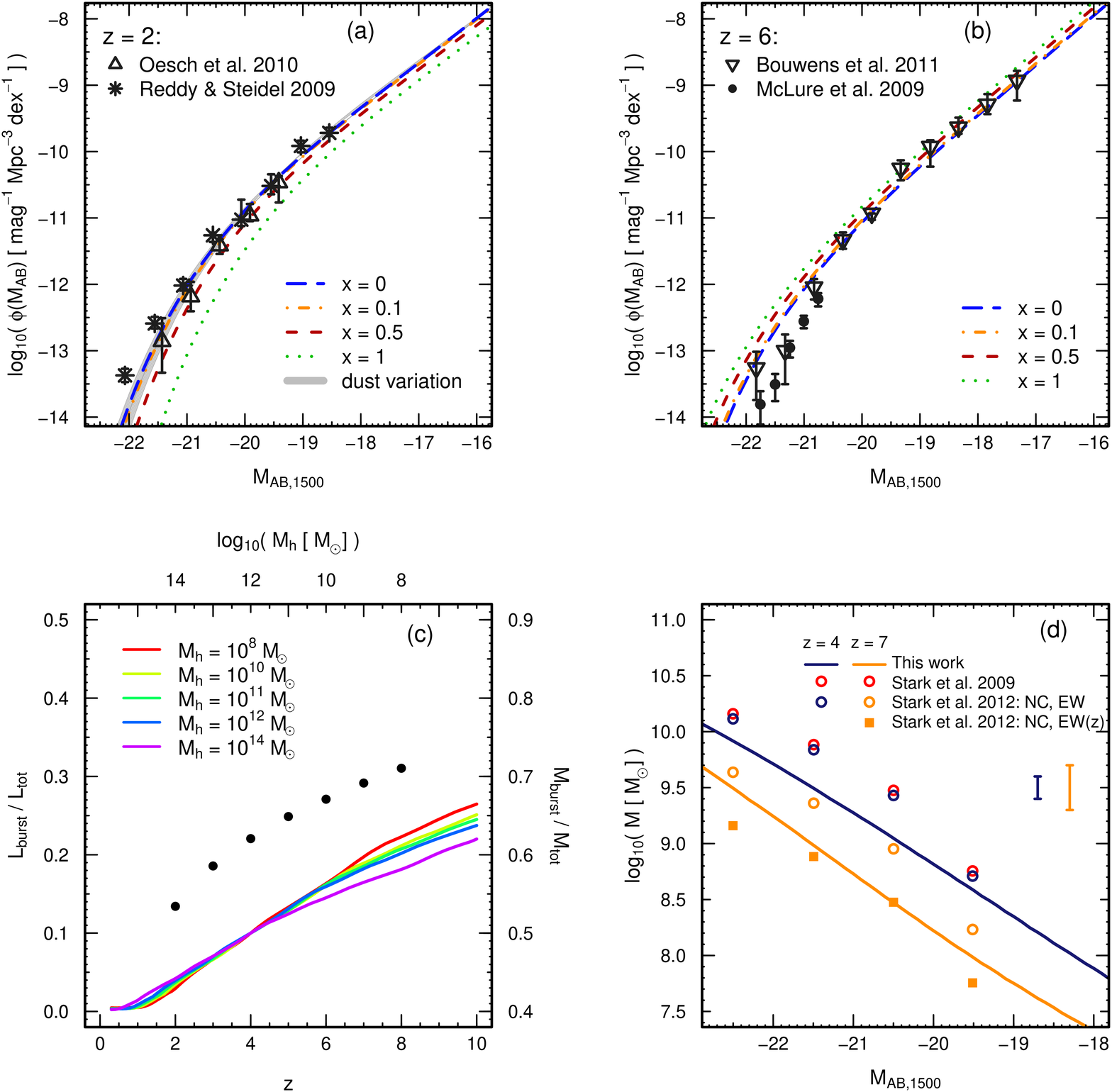} \caption{(a)-(b): Impact on the $z=2$ and $z=6$ LFs given by varying the relative contribution of burst to galaxy luminosity. Values $x\ll1$ give overall a good description of the observed LFs, while $x\sim1$ underestimates the LF at low $z$. Additionally in panel (a), the impact of variation of $\beta_{\mathrm{M_{UV}=-19.5}}$ by $\pm30\%$ is shown as gray area. (c): Contribution of the burst mode to the total UV luminosity as a function of redshift and halo mass (bottom-left axis) and to the total stellar mass as a function of halo mass, since there is no redshift dependence (top-right axis). (d): Stellar masses as a function of UV luminosity for $z\sim4$ and 7 (NC: nebular contamination corrected data with equivalent width evolution (EW($z$)) and without (EW), respectively).} \label{fig:sensitivity}\end{center}\end{figure*}
%%%%%%%%%%%%%%%%%%%%%%%%%%%%%%%%%%%%%%%%%%%%%

Figure~\ref{fig:sensitivity} (panels (a)-(b)) shows model-observations comparisons for the $z=2$ and 6 LFs; the model results are plotted for different values of the burst fractional contribution $x$ to the total  UV luminosity, and for different  dust extinction corrections (as parametrized by $\beta$). Varying  $\beta$ within the uncertainty   given by \citet{bouwens12a} has  little impact on the predicted LFs, which are thus robust against uncertainties in the amount and  treatment of dust obscuration.  In contrast, the predicted LFs do depend on the choice of  $x$. As the star formation histories are modeled to approach  a single episode of star formation at the halo assembly time, i.e., $x\rightarrow1$,  the $L(M_h,z)$ relation shows an  increasingly stronger dependence on the assembly time: at low redshift, halos become too faint relative to the observations, and the LFs and $\rho_L$ are under-estimated; at high redshifts, halos are too UV-bright and  LFs and $\rho_L$ are overestimated. 

By construction,  the initial burst phase contributes modestly to the UV luminosity of galaxies at all epochs, especially at lower redshifts, i.e., as $t_{age}$ increases. This is shown in panel (c) of Figure~\ref{fig:sensitivity}, which plots the contribution of the burst mode to the total galaxy luminosity, for halos of different masses. The redshift evolution of the $L_{burst}/L_{tot}$ ratio is faster for smaller halos, as for these $t_{age}$ evolves faster (Figure~\ref{fig:model}). In contrast, the initial star formation burst, occurring at the time when the halo has already  assembled half of its total mass,  consistently contributes of order a half of the stellar mass budget (Figure~\ref{fig:sensitivity}, panel (c)). Specifically, the contribution of the burst phase to the stellar mass is roughly 50\% at $M_{h}\sim10^{14}~\mathrm{M_\sun}$, and increases to about 70\% at $M_{h}\sim10^{9}~\mathrm{M_\sun}$.

Figure~\ref{fig:sensitivity}, panel (d), shows stellar mass as a function of magnitude. The red points show the relation at $z=4$ of \citet{stark09}, and the blue points are re-normalized for accounting for emission lines (see \citealt{stark13,de-barros12}). The model data are slightly below the $z=4$ observations; in contrast, at $z=7$ the model predicts a $M_{\star}-M_{AB,1500}$ relation which well matches the data of \citet{stark13}, once these are corrected for emission line contamination and for a redshift-dependent  equivalent width of nebular emission (increasing with increasing redshift).

The middle panel of Figure~\ref{fig:SFRD} shows the comparison between model and observations for $\rho_{M_{\star}}$. Overall, the model (with or without scatter in $L(M_h,z)$) fits the data well at all redshifts. At $z=7$, we under-estimate by 0.25 dex the (not emission-corrected) \citet{gonzalez11} data point. On the other hand, we are broadly consistent with the \citet{stark13} measurements. The sSFR is shown in the bottom panel of Figure~\ref{fig:SFRD} and shows overall a good agreement with the observations.

%%%%%%%%%%%%%%%%%%%%%%%%%%%%%%%%%
\section{Conclusion}\label{sec:Con}

We have presented a  model for the evolution of the UV LF based on the simple assumption that all massive DM halos host a galaxy with a star formation history  that is closely  related to the halo assembly time. We specifically adopt for the star formation histories  the combination of  an initial star formation burst at the halo assembly time, representing the integrated galaxy star formation histories down to this epoch, plus a constant SFR phase, a proxy for secular, low-level  star formation activity fueled by steady gas accretion. While the assumption that each DM halo hosts only one galaxy is clearly a simplification, especially towards lower redshifts, the model is remarkably successful in reproducing major features of the evolving star forming galaxy population since $z\simeq8$; this is also due to the fact that  massive galaxies which, at later times, will share a common halo, will be mostly quenched of their star formation activity. It is remarkable that this simple parametrization reproduces very well the  evolution of the UV LFs over the whole cosmic time since $z\simeq8$  down to $z\simeq0$, as well as  the evolution of the cosmic specific star formation rate, luminosity density and stellar mass density. In our model, the cosmic star formation rate density rises and then falls towards lower $z$ naturally as it is observed, which can be explained by the drop in the accretion rate at low redshifts of the individual DM halos - overtaking the increase in abundance of galaxies at $z\la2$. This  demonstrates the key role played by DM halo assembly in shaping the properties of the luminous galaxies.

%%%%%%%%%%%%%%%%%%%%%%%%%%%%%%%%% 

\end{document}